# Rubisco function, evolution, and engineering

Noam Prywes[1], Naiya R. Philips[2], Owen T. Tuck[3], Luis E. Valentin-Alvarado[4], David F. Savage[1,2,4,5,*]

[1]Innovative Genomics Institute, University of California, Berkeley, California 94720, USA;
[2]Department of Molecular and Cell Biology, University of California, Berkeley, California 94720, USA;
[3]Department of Chemistry, University of California, Berkeley, California 94720, USA;
[4]Graduate Group in Microbiology, University of California, Berkeley, California 94720, USA;
[5]Howard Hughes Medical Institute, University of California, Berkeley, California 94720, USA;
*Corresponding author, email: savage@berkeley.edu

## Abstract

Carbon fixation is the process by which $CO_2$ is converted from a gas into biomass. The Calvin Benson Bassham (CBB) cycle is the dominant carbon fixation pathway on earth, driving >99.5% (1) of the ~120 billion tons of carbon (2) that are "fixed" as sugar, by plants, algae and cyanobacteria. The carboxylase enzyme in the CBB, ribulose-1,5-bisphosphate carboxylase/oxygenase (rubisco), fixes one $CO_2$ molecule per turn of the cycle. Despite being critical to the assimilation of carbon, rubisco's kinetic rate is not very fast (3) and it is a bottleneck in flux through the pathway (4). This presents a paradox - why hasn't rubisco evolved to be a better catalyst? Many hypothesize that the catalytic mechanism of rubisco is subject to one or more trade-offs(5), and that rubisco variants have been optimized for their native physiological environment (6). Here we review the evolution and biochemistry of rubisco through the lens of structure and mechanism in order to understand what trade-offs limit its improvement. We also review the many attempts to improve rubisco itself and, thereby, promote plant growth.

## Introduction

Rubisco is the carboxylase of the Calvin-Benson-Bassham cycle (CBB), where it fixes $CO_2$ onto a ribulose bisphosphate (RuBP) sugar, producing two molecules of 3-phosphoglycerate (3PG). This reaction is central to the Carbon Cycle and converts ~100 gigatons of carbon from $CO_2$ into biomass annually — approximately 10x more than annual human emissions.

The balance of $CO_2$ uptake and release by the biosphere has supported a stable atmospheric composition over the past few millions years of ~78% $N_2$, ~21% $O_2$, and ≈0.02-0.04% $CO_2$, but this was not always the case. Some ~3 billion years ago, when rubisco first evolved, $CO_2$ levels were likely quite high and $O_2$ certainly quite low (7). The historical atmospheric composition may help explain one of the seeming paradoxes of rubisco biochemistry: rubisco is notable among carboxylases for also reacting promiscuously with $O_2$ (Table 1). The off-target oxygenation of RuBP produces 3PG and a molecule of 2-phosphoglycolate (2PG), the latter of which must be recycled in a carbon salvage pathway, termed photorespiration. Ironically, the oxygenic photosynthetic metabolism enabled by rubisco caused the rise in atmospheric $O_2$ levels and the attendant inhibition of the enzyme.

The apparent inefficiency of rubisco has led some to malign it as "slow" or "confused" (8) by some and "average" (3) or "not really so bad" (9). In terms of maximum rate ($k_{cat}$) or catalytic efficiency ($k_{cat}/K_M$), rubisco ranks close to the median among characterized enzymes (3). It is far from a diffusion-limited "perfect" enzyme, but perhaps rubisco deserves to be graded on a curve: "perfect" enzymes often catalyze "easier" reactions — reactions that proceed at appreciable uncatalyzed rates. The uncatalyzed carboxylation of RuBP is far too slow to measure, but its rate constants have been estimated computationally (10). By this measure, rubisco is highly effective, conferring a $10^{15}$ to $10^{18}$-fold rate enhancement (9). This is a reflection of the difficulty of rubisco's chemical mechanism, in particular the challenge associated with having $CO_2$ — a small, hydrophobic, uncharged, relatively inert molecule — as a substrate. As a result of its kinetic parameters, rubisco is highly expressed in plants to increase total carboxylation and is therefore the most abundant protein on the planet (2).

Rubisco has received significant attention as a target for protein engineering, but attempts to improve it face a steeper challenge than is typical (11, 12). Successful protein engineering campaigns often optimize a property orthogonal to the natural function of an enzyme; for example, one starts with a small amount of promiscuous activity in an enzyme that catalyzes a different reaction and, over successive rounds of screening or selection, cultivates the promiscuous activity (13). In contrast to these empirical best practices, the desired improvements to rubisco (carboxylation rate, $CO_2$ affinity, and specificity for $CO_2$ over $O_2$) are all axes along which evolution has already acted for billions of years with apparently limited results (14). Improvements in one rubisco biochemical parameter (e.g. kinetic rate of carboxylation $k_{cat,C}$) may come at the expense of another (e.g. specificity $S_{C/O}$). A variety of trade-offs, discussed below, have been proposed based on biochemical data and mechanistic chemical logic (5, 6). Relevant parameters include rate constants $k_{cat,C}$ and $k_{cat,O}$; Michaelis constants $K_C$ and $K_O$; and the ratio of carbon and oxygen-specific catalytic efficiencies $S_{C/O}$.

The earliest rubisco mutagenesis studies were motivated by a desire to improve rubisco's rate and specificity (15, 16). Efforts have focused on regions near the active site and a mobile region termed loop 6 that is known to be involved in catalysis (17). Rational mutagenesis has not been successful in generating superlative rubiscos, i.e. enhancing carboxylation while limiting oxygenation. Mutant libraries in bacteria have been used to evolve rubisco in high throughput since the early 1990s (18). *E. coli* has been engineered in a variety of ways to serve as a chassis for rubisco library selections (19). These selections have generally resulted in rubiscos with improved expression and stability, though some have yielded enzymes with faster $k_{cat,C}$s (20, 21) and lower $K_C$s (22). An alternative means of understanding the sequence-function landscape is through natural diversity. Although initial studies focused on plant rubisco, the bulk of sequence and functional diversity is found in microbes (Davidi et al. 2020; Tabita et al. 2008) and a wide swath of this diversity remains unexplored.

Even if rubisco could be improved, it is debated whether this would translate to faster growing autotrophs. At low light or high $CO_2$ concentrations, the maximum rate of electron transport ($J_{max}$) limits growth (23, 24). A similar argument has been made regarding improvements to specificity. Although photorespiration may seem wasteful, it may be necessary for nitrogen uptake (25, 26), and nitrogen is often the limiting factor for growth (27). However, increased rubisco carboxylation flux may improve growth and yield in agricultural settings (28, 29), when resources other than carbon are not limiting (30), or when the downstream carbon sink capacity is artificially increased (31). Thus far, however, all attempts to improve plant growth by rubisco replacement have failed.

Here we review recent and historical insights into one of the most pivotal enzymes to life on earth. We review how the structure and catalytic mechanism influence the catalytic rate of rubisco and then explore the diverse superfamily of rubisco homologs. Finally, we summarize the history of attempts to improve rubisco function both *in vitro* and *in vivo* and finish by highlighting where new insights and practical advances may be achieved.

## Box 1: Fraction 1 Protein - A Brief History

Rubisco was discovered twice, first as an unidentified enzyme of unusually high abundance in leaves and later as the first step of $CO_2$ assimilation in plants. Although the initial discovery, that as much as 50% of soluble leaf protein is a single species in gel electrophoresis, was viewed skeptically, repeated crystallization (32) would prove that there was only one protein in the suspect gel band. Originally called "Fraction 1 protein" based on ammonium sulfate fractionation (33), it was eventually realized this protein was the same enzyme identified as the first step of the CBB pathway — ribulose diphosphate carboxylase (34, 35). C3 photosynthesis is named after phosphoglycerate, the triose product of the carboxylase reaction. Calvin and coworkers had already correctly speculated that there was an enediolate intermediate (Figure 2B, center) bound to magnesium which would attack $CO_2$ and subsequently hydrolyze (36). The relatively slow activity of rubisco was also noted with curiosity in these early studies (37). Different forms of rubisco were first recognized in *R. rubrum* and *Rhodobacter* in 1968 (38).

It took until the early 1970s to learn that rubisco engaged in promiscuous oxygenation activity (39, 40), which led to its current name — ribulose bisphosphate carboxylase/oxygenase — provided by David Eisenberg (33). Careful analysis of the order of addition of reactants further revealed that rubisco requires an “activation” step wherein a specific active site lysine (41, 42) is carbamylated with $CO_2$ (43, 44) — there are thus 2 $CO_2$ molecules at the active site during catalysis, one as a cofactor and the other as a substrate. Carbonic anhydrase analysis showed that $CO_2$ is the substrate for carboxylation and not bicarbonate (45), tritium exchange experiments proved Calvin was right about reversible enediol formation (46, 47) and NMR studies confirmed the order of cofactor, metal and substrate binding (48). Mendelian inheritance of the small subunit (49) and maternal inheritance of the large subunit (50) demonstrated the respective nuclear and chloroplastic localization of the genes. Just a few years after the first DNA sequence of maize rubisco was determined (51), mutagenesis studies were pinpointing the key active site residues conserved between bacterial and plant rubiscos (52, 53). At the same time, x-ray crystal structures were solved for both of these forms (54, 55). For more history see (33, 56).

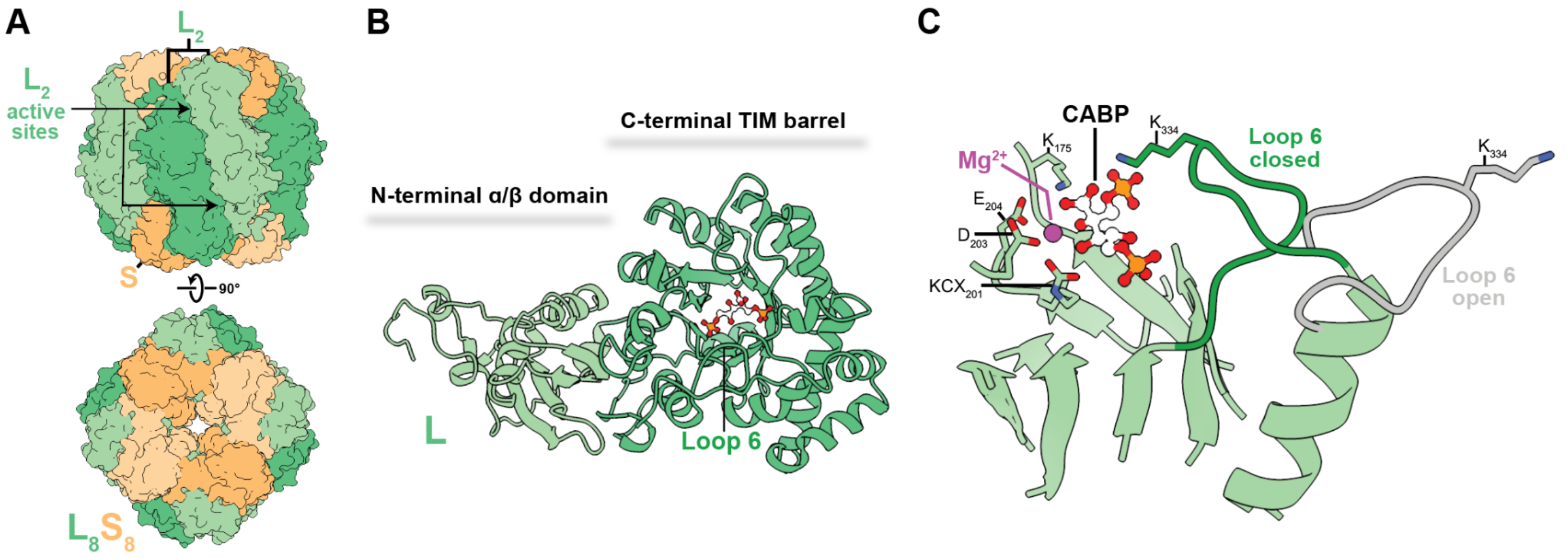

**Figure 1.**
A triptych digest of structural features of Form I rubisco from spinach bound to inhibitor 2-carboxyarabinitol-1,5-bisphosphate (CABP) (PDB IDs: 8RUC, 9RUB, (57, 58). **A)** The $L_8S_8$ architecture contains the conserved $L_2$ homodimeric functional unit present in all rubiscos. **B)** Substrate binds at the face of the C-terminal TIM barrel after capture of divalent cation and carbamylation of an active site lysine. CABP mimics the structure of RuBP after $CO_2$ addition. **C)** Loop 6 extends over the center of the TIM-barrel during binding and participates in catalysis (green, +CABP; gray, apo conformation).

# Structure

All rubiscos have the same basic functional unit of a homodimer of large subunits ($L_2$), with each subunit comprising an N-terminal α/β domain and C-terminal $(β/α)_8$ TIM barrel (Figure 1A,B). The structure of the rubisco large subunit dimer is well-characterized and highly conserved, even between homologs with only ~30% sequence identity (59, 60). The subunits are antiparallel with the two active sites situated at the interfaces between the N- and C-termini. A polar binding pocket is formed across the opening of one C-terminal barrel with positively charged regions on opposite sides of the barrel to anchor the two phosphate groups on RuBP while the coordinated $Mg^{2+}$ orients C2 of the substrate for its reaction with $CO_2$ (Figure 1B,C, 2A-C). This pocket is formed primarily by highly-conserved residues in the loops connecting the β strands and α helices of the barrel, while two loop regions from the neighboring N-terminal domain close part of the top of the binding pocket (61). Upon binding to RuBP, the flexible loop 6 connecting β-strand 6 to α-helix 6 in the C-terminal domain undergoes a ≈12 Å shift that encloses the active site (62, 63), and is thought to stabilize the reaction intermediates. All *bona fide* rubiscos have a high degree of conservation of the active site residues (64), and mutations to active site residues have been reviewed in detail (65, 66). Mutation to any of the roughly 20 conserved residues surrounding the active site, discussed more below, leads to a decrease in catalytic activity or specificity.

The rubisco family of proteins contains a number of forms, separated by sequence, taxonomic distribution, or biochemical characteristics (see Evolution below). The $L_2$ dimer is the core feature of rubisco structure, but it often assembles into high order oligomers across the phylogeny. For example, the first rubisco that was isolated (67) was found to assemble into a heterohexadecameric structure ($L_8S_8$) composed of four $L_2$ dimers assembled in a ring, with the dimer:dimer interfaces capped at the top and bottom by additional small subunits (SSUs) (68, 69) (Figure 1A). This class of rubisco is called Form I, and has been further classified into subtypes IA, found mostly in proteobacteria, cyanobacteria with α-carboxysomes and the alga Paulinella; IB in β-cyanobacteria, green algae and plants; IC mostly in proteobacteria; ID in red algae, diatoms, haptophytes and other algae; and IE in actinobacteria and a variety of newly characterized and unknown forms in Chloroflexi bacteria (Figure 4C) (60, 70). Form II, II/III and III rubiscos lack the small subunit but often form higher-order oligomers of the $L_2$ dimer. They are primarily found in prokaryotes.

The SSU is not strictly necessary for carboxylation, but its absence results in a dramatic loss of complex stability, substrate binding affinity, and catalytic activity in model plant rubiscos (71), leading to some debate over its function (reviewed in (72)). Form I is the only class of rubiscos with an SSU and also demonstrates a higher $CO_2$-specificity than other forms, suggesting the SSU plays some role in selectivity. However, there is currently no confirmed mechanism by which the SSU provides this increased specificity. One theory is that the SSU stabilizes the large subunit, improving its robustness to mutation and allowing it to explore a productive region of the fitness landscape with higher $CO_2$-specificity (73). Removing the SSU greatly reduces the thermal stability of the *Synechococcus elongatus* 6301 heterohexadecamer, but it isn't absolutely required for the formation of stable oligomers, as Form I' rubiscos form octamers and Forms II and III form $(L_2)_n$ assemblies without SSUs (74, 75). The SSU has more recently been shown to have a possible role in regulating rubisco activity, as differentially expressed isoforms of the tobacco rubisco SSUs yield holoenzymes with moderate differences in $k_{cat,C}$ and $K_C$ values (76). It has also been suggested that the formation of the holoenzyme creates unique binding sites at the subunit interfaces wherein rubisco can serve as a protein-protein interaction hub to recruit additional factors (77–80). It is also worth noting that fusing the large and small subunits into one protein produces functional rubisco in tobacco (81), which will aid in further research efforts to determine the role of individual SSU isoforms *in planta*.

While Form I rubiscos tend to have higher specificity for $CO_2$ and more opportunity for regulation, they also require a suite of folding chaperones in order to avoid misfolding and aggregation (reviewed in (82)). *Arabidopsis thaliana* rubisco has been shown to require five chloroplast-specific chaperones as well as the GroEL/ES chaperonins to fold in *E. coli*, and these chaperones were species-specific. That is, the *E. coli* strain expressing *A. thaliana* rubisco chaperones could only produce *A. thaliana* rubisco and not *N. tabacum* rubisco (83). This presents a major challenge to studying Form I rubiscos: heterologous expression of a given rubisco requires first identification and expression of its accompanying chaperones, which is itself a tedious process in slow-growing plants or unculturable bacteria and algae.

## Mechanisms

Rubisco catalysis occurs in four composite steps: 1) preparation of the active site, 2) RuBP binding and enediol formation, 3) gas addition ($CO_2$ or $O_2$) and 4) cleavage to produce two 3PG or one 3PG and one 2PG, respectively (Figure 2A)**.** Computational modeling of rubisco led by Gready and Tcherkez (84, 85) permits speculation about the heights of the energy barriers (Figure 2B). Here, we summarize key mechanistic steps and current proposals (Figure 2C-E). Efforts to improve rubisco activity will benefit from considering the nuances of these individual steps along with the chemical and structural constraints imposed by the active site.

Carbamylation of a conserved active site lysine is the crucial activation step required for both carboxylation and oxygenation (Figure 2C). The carbamyl moiety is derived from the attack of a lysyl amine on a $CO_2$ molecule (a different $CO_2$ molecule than the one which will be added to RuBP), generating a zwitterionic intermediate. The coordinated divalent $Mg^{2+}$ stabilizes the *aci* form of the carbamate (State 3, Figure 2C), prolonging the modification and priming the carbamyl group for its function as a general base in later stages of catalysis (41, 43, 86).

After $Mg^{2+}$ binding and carbamylation, RuBP enters the active site, binds to $Mg^{2+}$ (State 4, Figure 2C), and is converted to the *cis*-2,3-enediolate, with the carbamyl oxygen that is not bound to $Mg^{2+}$ acting as a general base (Reviewed extensively in (86)). Evidence from crystallography and isotopic labeling suggest that proton transfer activates the enediolate such that the enediolate oxygens are nearly bidentate with respect to the metal (State 6, Figure 2C). This results in a strained pseudo-five-membered ring poised to act as a nucleophile. In carboxylation, $CO_2$ approaches from the solvent-exposed face and adds along with $H_2O$ across the enediolate to give the 6-carbon intermediate. Rubisco does not bind $CO_2$ non-covalently with a dedicated binding pocket, in a classical Michaelis complex. Rather, covalent bond formation happens immediately upon $CO_2$ entering the active site (48). There is debate over the mechanism of hydration and $CO_2$ addition: Lorimer and colleagues showed that the mechanism was most likely concerted, with His294 acting as the base (87) (Figure 3C). Additional evidence from chemical simulations (88) and isotope experiments (89) complicate these assertions, though no conclusive mechanism can be drawn to-date. The structure of rubisco bound to the competitive inhibitor carboxyarabinitol-1,5-bisphosphate (CABP), a mimic of the reaction intermediate found in State 8 in Figure 2D, illustrates the location of the nascent carboxyl group and serves as a useful tool for dissecting the mechanism. Once hydrated (State 8, Figure 2D), cleavage of the long C2-C3 bond occurs rapidly via carbamate-mediated deprotonation (86) and stereospecific protonation, furnishing two molecules of 3PG bound for central metabolism (Figure 4B).

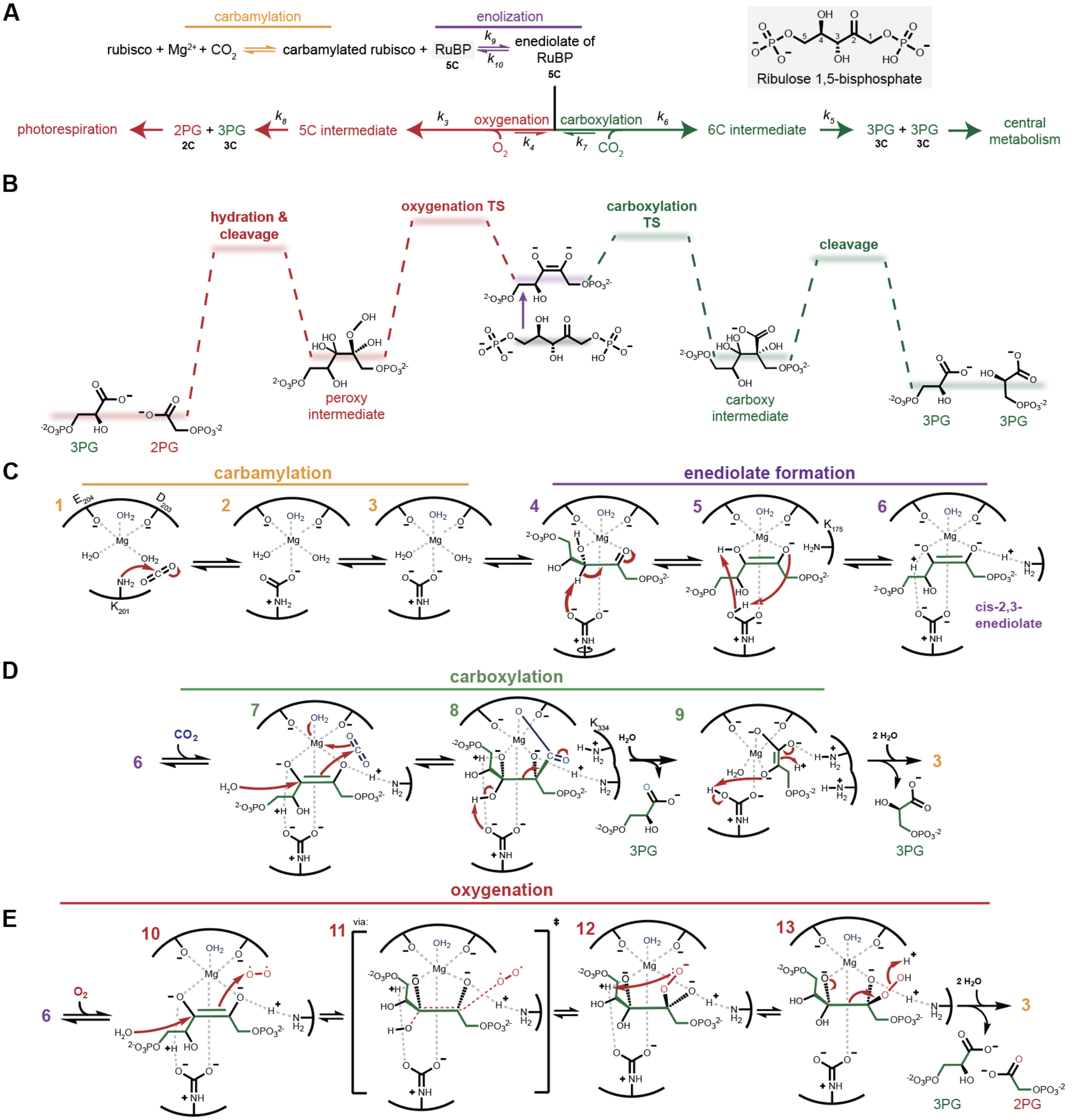

**Figure 2.**

Mechanisms of rubisco catalysis. **A)** Middle-out scheme, rate constants are numbered as in (5, 90). **B)** Simplified landscape of chemical reactions at the rubisco active site. Relative intermediate and transition state energies are not drawn to scale. **C)** Carbamylation, RuBP binding, and preparation of the activated *cis*-2,3-enediolate **6** (86). **D)** Carboxylation of enediolate intermediate **6**. Gas addition and hydration are shown as a concerted mechanistic step. **E)** Oxygenation of **6**, giving one molecule of 3PG and one molecule of 2PG. Drawn to depict plausible steps in single electron transfer (SET). Proton assignments informed by isotope labeling experiments in (91). All active site residues are numbered based on spinach rubisco. RuBP: ribulose bisphosphate, 2PG: 2-phosphoglycolate, 3PG: 3-phosphoglycerate, TS: transition state.

Rubisco oxygenation, which produces one 3PG and one photorespiration-bound 2PG, is notable because the reaction of the singlet endiolate with the biradical triplet oxygen is spin-forbidden and, in nearly all other oxygenases besides rubisco, is achieved with cofactors. Indeed, no other carboxylase has promiscuous oxygenase activity (Table 1). Experimentalists have only recently begun to elucidate the mechanism of rubisco oxygenation. Work by Tcherkez and colleagues demonstrated that rubisco catalyzes oxygenation by single electron transfer (SET), wherein a superoxide radical transfers an electron to RuBP generating two radicals which can recombine freely. They base their argument on both the invariance of the reaction to the $^{18}O$ kinetic isotope effect in rubiscos with different $CO_2/O_2$ specificities, and computed reduction potentials of RuBP and superoxide (91). These data suggest that the enediolate is energetically poised for SET oxygenation and that diversity in $S_{C/O}$ is the result of minute differences in active site electric environments of various rubiscos (92). Another plausible mechanism for spin-forbidden oxygenation is intersystem crossing (ISC) where one reactant changes its electronic state to match the other. For example, torsionally strained RuBP in the active site can, in principle, be excited from a singlet to a triplet state. Kannappan, Cummins, and Gready performed large-scale quantum chemical calculations which suggest that, in disagreement with the SET model, the enediolate forms a caged biradical complex with $O_2$ (84). Both the ISC and SET camps agree that preparation of the enediolate for facile carboxylation invites oxygenation, and that there is no obvious way to rationally design a rubisco whose active site configuration strongly disfavors oxygen addition without compromising carboxylation. The discovery of an oxygenase-only rubisco isoform (*vide infra*) suggests that the relationship between oxygenation and carboxylation may not be symmetric, carboxylation may necessitate oxygenation, but the reverse is not the case: it is possible to arrange an active site with an apparent $S_{C/O}$ of zero (93). This result is less surprising when we consider that rubisco carboxylation, while favorable energetically, with a $\Delta G^\circ \approx -30$ kJ/mol, pales in comparison to the oxygenation $\Delta G^\circ \approx -500$ kJ/mol. Similarly, it was shown that carboxylation can also be suppressed by replacing $Mg^{2+}$ with $Co^{2+}$ in the active site (94).

Several post-translational modifications apart from lysine carbamylation can interface with the rubisco mechanism. Nitrosylation of an active site cysteine occludes RuBP from binding before carbamylation is complete (95, 96). RuBP binds extremely strongly even in uncarbamylated active sites, impeding activity and requiring a chaperone, rubisco activase, to remove RuBP and regenerate the active site (97) (reviewed in (82)). Rubisco S-nitrosylation via nitric oxide (NO) is increasingly appreciated as an important modulator in plant metabolism and may represent a potential avenue for control of rubisco activity (98). Some autotrophs exert additional control over the active site by using binders that mimic catalytic intermediates. For example, the dephosphorylated analog of the CKABP intermediate (Figure 2B,D) carboxyarabinitol 1-phosphate (CA1P) binds the carbamylated active site in the absence of sulfate ions, and has been implicated in protection of rubisco from proteolysis during night-time inactivity (99, 100).

## Kinetics, trade-offs, and optimization

It has long been observed that specific rubiscos (i.e. those with high $S_{C/O}$) tend to be slower carboxylases (i.e. have low $k_{cat,C}$) (101). The characterization of natural and engineered rubisco variants led to the emergence of several trade-off models to explain such correlations in the biochemical data. Kinetic models for rubisco activity have been developed *in vitro* (5, 90, 102), and *in vivo* (103) though only a few of the many possible parameters are routinely measured. As kinetic measurements of rubiscos from various species trickled in over the decades, a picture emerged of an enzyme caught between two or more conflicting priorities, forced to compromise (5, 6, 101). While there are still very few available measurements from many rubisco forms (Figure S2), it is possible to compare kinetics of rubiscos from individual classes, especially the Form IB variety found in many cyanobacteria, green algae, and all plant chloroplasts, which are the most extensively studied.

One clear trend in kinetic parameters across species is the correlation between catalytic efficiencies for carboxylation and oxygenation (i.e. $k_{cat,C}/K_C$ and $k_{cat,O}/K_O$) (Figure 3). This result has been interpreted to imply that while rubisco can easily evolve to increase or decrease both carboxylation and oxygenation efficiencies, it is very difficult to improve the ratio between them (6, 14). Since the selectivity measure is simply the ratio between carboxylation and oxygenation efficiencies ($S_{C/O} = k_{cat,C}/K_C$ / $(k_{cat,O}/K_O)$), it follows that $S_{C/O}$ should not vary greatly (Figure 3A). Within the Form IB rubiscos, $S_{C/O}$ varies only 30% (14, 104). A related trade-off, between selectivity and rate, is common among enzymes, and many enzymes, like rubisco, have a hard time distinguishing between substrates through different binding energies. These enzymes are forced to use transition state stabilization differences (11). Mechanistically, the observed trade-off may be related to initial enediolate formation which can then react with $O_2$ or $CO_2$ with little ability to discriminate (Figure 3B). The correlations supporting other proposed trade-offs, like that between $k_{cat,C}$ and $K_C$ (5, 6) have attenuated as more data have been collected (14, 104, 105).

$k_{cat}$ and $K_M$ are composites of the rate constants for individual steps in the reaction mechanism, so correlations between these phenomenological parameters must reflect dependencies between the energy barriers associated with individual reaction steps. For instance, it has been proposed that the observed correlation between carboxylation and oxygenation efficiencies can be rewritten as a correlation between the on-rate of $CO_2$ and $O_2$ (i.e. $k_3$ is proportional to $k_6$, Figure 3B). This correlation in turn can be explained by the proposal that the activation energies for carboxylation and oxygenation can grow or shrink but only in proportion to one another. A simple way for this to happen would be if the energy of the enediolate were to change while the transition state energies remain fixed (Figure 3B) (6, 14, 85, 106). As the energy of the bound enediolate rises, the Hammond postulate dictates that the transition states become more reactant-like, which would result in worse selectivity because the reactant in carboxylation and oxygenation is the same (5). Finally, the basis for a trade-off can ultimately be found in the structure and electronics of the enzyme active site. One proposal for the modest negative correlation between $k_{cat,C}$ and $K_C$ is that a key active-site residue, Lys175, may act as a base in enolization and as a acid in hydration/cleavage, which could result in a trade-off where improvement to one of its functions comes at a cost to the other. Other possible trade-off sources are shown in Figure 3C (107, 108). Advances in molecular modeling (88), more measurements of individual rate constants from different species (102), more data on rubiscos outside of Form I (109) and molecular evolution studies (*vide infra*) will improve our understanding of the mechanistic basis, or lack thereof, of apparent trade-offs.

A final trend relates to host physiology and strengthens the case for trade-offs. Many organisms possess $CO_2$ concentrating mechanisms (CCMs) which increase the concentration of $CO_2$ near rubisco (110, 111). Elevated $CO_2$ promotes carboxylation and competitively inhibits oxygenation (112). CCMs are found in plants, algae and bacteria and have evolved convergently resulting in a variety of different strategies. In bacteria, CCMs sequester rubisco inside a protein compartment with locally high $CO_2$ produced via active carbon uptake. Alternatively, so-called C4 plants use specialized anatomy to separate initial $CO_2$ fixation with a non-rubisco carboxylase (e.g. phosphoenolpyruvate carboxylase) in order to facilitate a second rubisco carboxylation, the latter carried out in specialized cells with high $CO_2$ (113). Rubiscos from CCM-bearing organisms generally have higher carboxylation rate constants ($k_{cat,C}$), lower selectivities ($S_{C/O}$) and weaker affinities for $CO_2$ ($K_C$) (6, 14, 104), implying that evolution cannot isolate a superlative variant and instead optimizes the system around rubisco to achieve fast, on-target catalysis.

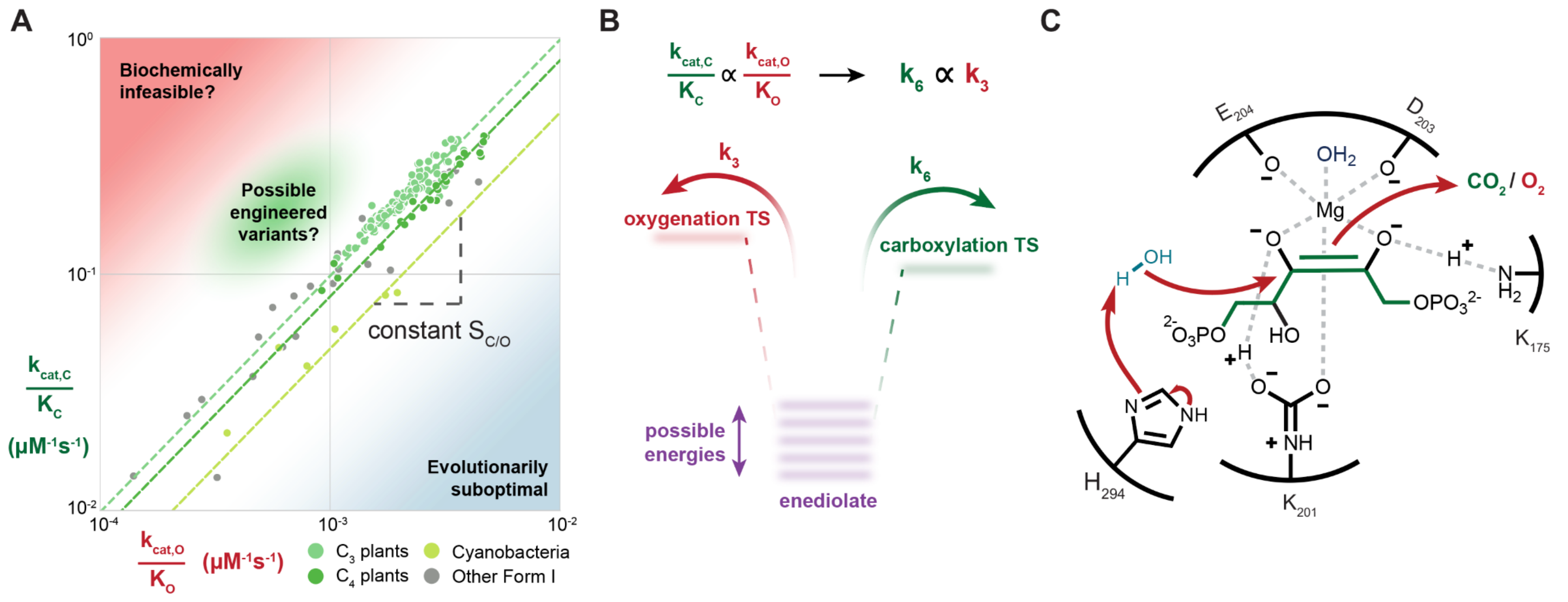

**Figure 3:**
Example of a possible mechanistic trade-off - the empirical link between carboxylation and oxygenation efficiencies. **A)** Catalytic efficiencies for carboxylation and oxygenation for Form I rubiscos (data from (14)). The values are fit to a line in log-space that has a slope very close to 1. **B)** The trade-off represented by the line between measured kinetic parameters can be interpreted as a correlation between the rate constants for gas addition of $CO_2$ and $O_2$. The correlation could arise from a change in the energy of the enediolate bound in the active site with the transition state energies held constant. **C)** More detail for State 6 in Figure 1C. Some of the residues that may be involved in determining the energy of bound enediolate and the activation energies for gas addition are depicted. This step of the mechanism precedes both oxygenation and carboxylation.

# Box 2: Carboxylases other than rubisco

When considering the complex mechanism of rubisco catalysis, and the steps peculiar to the reaction(s) it catalyzes, a paradox begins to materialize: Why is >99% (1) of carbon still fixed through the CBB cycle despite rubisco's well-known limitations? This question follows from an intuitive premise that evolution could surely do better than relying on a slow and non-specific catalyst to produce nearly all organic carbon in the biosphere. One place to look for answers is among other carboxylases — there are dozens of carboxylases and roughly ten different carbon fixation pathways (114, 115). These comparisons offer three primary answers, none of them completely satisfying: Oxygen sensitivity, energetic cost, and metabolic inflexibility. Some pathways rely on enzymes that are non-functional, or nearly so, in the presence of oxygen (e.g. reverse TCA cycle, see (116)). Oxygen insensitive pathways generally require more energy per fixed carbon. Finally, the Calvin cycle is well-integrated into the pentose phosphate pathway and the remainder of central carbon metabolism (Figure S3), while the two alternative oxygen-tolerant pathways found in prokaryotes (114, 115) use metabolites that are less integrated into central metabolism.

Various carboxylases, and decarboxylases run in reverse, have been engineered to produce organic molecules using $CO_2$ as a substrate (117). Decarboxylases often display worse thermodynamics and kinetics than carboxylases but in many cases can be forced to catalyze carboxylation especially in the presence of elevated $CO_2$ (118, 119). Among carboxylases, rubisco is unique in several respects:

1) It is the only one to carboxylate RuBP and thus uniquely compatible with the CBB cycle.
2) Among carboxylases, it is the only one that reacts promiscuously with $O_2$.
3) It is in a small minority of carboxylases that do not directly couple activity to an energetic cofactor (other CBB reactions supply the required energy and reducing potential).

Some clear advantages exist among alternative carboxylases:

1) Many carboxylases use the more accessible substrate bicarbonate, which is easier to bind due to its charge and high concentration (typically ~100x the concentration of $CO_2$ in water at intracellular pH). These enzymes gain the ability to separate substrate binding from reactivity by using an ATP to convert bicarbonate to carboxyphosphate in a classic umpolung mechanism wherein a nucleophile becomes an electrophile (120).
2) Some enzymes reduce their substrates concomitant with fixation, which typically reduces the number of ATPs hydrolyzed in their pathways overall, improving energetic efficiency (121).

Despite rubisco's drawbacks, it is unlikely that a scalable, engineered pathway will dethrone the CBB in the near future. Many carboxylases would be difficult to integrate into carbon fixation pathways because their substrates and products are not part of central metabolism (e.g. the aromatic carboxylases, Table 1). Alternative synthetic carbon fixation cycles using other carboxylases have also been proposed (122–125) and engineering these cycles *in vivo* is an ongoing grand challenge for synthetic biology. A variant of the CBB, on the other hand, has been successfully engineered into *E. coli* strains, converting them from heterotrophs to autotrophs *(*126, 127*)*. This was possible, in part, because *E. coli* lacks only two CBB enzymes — *prk* and rubisco — the remainder being part of the ubiquitous pentose phosphate pathway. As synthetic biology approaches to metabolic engineering become more widespread, extensive pathway engineering and host integration may open the door to replacing the CBB in microbes and eventually in plants.

| **Carboxylase Class** | **Difficulty with oxygen?** | **Energetic coupling?** | **Carbon fixation pathway** | **Substrate** |
|---|---|---|---|---|
| Rubisco | Confuses $O_2$ and $CO_2$ | Substrate (C-C cleavage) | Natural and artificial | $CO_2$ |
| Reducing carboxylases (e.g. crotonyl-CoA carboxylase) | None | NAD(P)H | Artificial | $CO_2$ |
| ATP-dependent carboxylases (e.g. Ac-CoA carboxylase) | None | ATP | Natural and artificial | $HCO_3$ |
| Vitamin K-dependent carboxylase | None | $O_2$ | No | $CO_2$ |
| Ferredoxin oxidoreductases (e.g. KGOR) | Usually $O_2$ intolerant | Ferredoxin | Natural and artificial | $CO_2$ |
| Amine carboxylases (e.g. carbamoyl phosphate synthase) | None | ATP | No | $HCO_3$ |
| Glycine dehydrogenase | None | Reduced disulfide | No | $CO_2$ |
| Aromatic carboxylase (PurE or PurK) | None | Substrate or ATP | No | $HCO_3$ |
| Formate dehydrogenase* | Usually $O_2$ intolerant | NAD(P)H, Ferredoxin or Ferrocytochrome b1 | Natural | $CO_2$ |

* While not technically carboxylases, formate (and carbon monoxide) dehydrogenases reduce $CO_2$ for use in metabolism.

# Evolution

After the initial discovery of a distinction between Form I $L_8S_8$ plant rubiscos and Form II $L_2$ microbial rubiscos (38), additional forms have been discovered. Form III rubiscos were discovered in 1999 (128, 129);

they are distinguished from other forms by their presence in archaea and their association with non-CBB pathways. They account for a large proportion of the diversity of rubiscos and many do not fall neatly into well-defined clades (Figure 4C). There is no formal threshold for the establishment of a new form and the phylogeny and nomenclature of these proteins remains in flux.

Rubisco enzymes are not always associated with the CBB cycle (130). Some plants use rubisco outside of the CBB to improve the efficiency of seed oil biosynthesis (131). In some heterotrophic bacteria the CBB can serve as a secondary electron sink (130). Form II/III and III rubiscos typically serve roles in nucleoside salvage (132, 133) or the reductive hexulose-phosphate pathway (134) and not in any version of the CBB. They are found in genomes apparently lacking phosphoribulokinase (*prk*) genes encoding the enzyme immediately upstream of rubisco in the CBB cycle. However, exceptions have been found in some Gottesmanbacteria, Deltaproteobacteria, and Chloroflexi species; they have *prk* in their genomes and are proposed to operate some modified form of the CBB (135, 136). In our phylogenetic analysis, we find that the latter group of rubiscos branch adjacent to the most divergent Form I rubiscos (compare Form III-transaldolase, the asterisk, to Form Iα in Figure 4C), inviting evolutionary speculation that would benefit from additional structural studies and metagenomic sequencing.

Form IV rubiscos were discovered in 2000 (137) and have alternately been called rubisco-like proteins (RLP; hereafter we will use “rubisco” as a term that excludes RLPs) because they catalyze reactions other than RuBP carboxylation. Due to their clear homology to rubiscos, RLPs were immediately recognized as either evolutionary precursors or derivatives (137). A number of different chemistries have been identified among various RLP clades, and one clade is known to harbor RLPs that catalyze at least 3 chemical reactions (Figure 4C). Despite recent discoveries of RLP chemistries (93, 138), the majority of named clades have no known function and some clades may be diverse enough to harbor additional chemistries (e.g. DeepYkr). Some RLPs can complement knockouts of others (139), as can *bona fide* rubiscos (140). Thus, despite extreme divergence in sequence, the underlying mechanisms of the various RLPs and rubisco are similar enough to catalyze each others’ reactions. It is unknown if this plasticity is a contributor to trade-offs in rubisco catalysis.

Sequences from environmental samples have repeatedly shed light on the evolutionary path of rubisco. The observation that the rubisco tree has many deeply branching archaeal Form III rubiscos led to the hypothesis that rubisco evolution begins here and, through many instances of horizontal gene transfer and structural innovations, led to the phylogenetic distribution observed today (60). By this hypothesis, the ancestor was a *bona fide* rubisco (Figure 4B). An alternative hypothesis, based on mechanistic considerations and structural parsimony, is that *bona fide* rubiscos evolved from an ancestral RLP (141). Rubiscos and all RLPs use mechanisms that involve the formation of an enolate (Table S3), but only the mechanism of rubisco contains a second step where the enolate attacks $CO_2$ - RLP enolates usually attack protons (141). However, one RLP does have a mechanism with a hydrolysis step, like rubisco (138)), and another exclusively catalyzes oxygenation of the same intermediate as rubisco - bending the definition of RLPs (93). The discovery of a suitable TIM-barrel protein to act as an outgroup to the rubisco superfamily could settle this debate (Figure 4B).

More recent evolutionary developments have been more amenable to phylogenetic analysis. Horizontal gene transfer (HGT) of rubiscos is widespread (64, 136). Plants acquired their Form IB rubiscos from cyanobacteria during endosymbiosis - with subsequent partitioning of the LSU and SSU between the chloroplast and nuclear genomes, respectively. Plant SSUs evolve faster than the LSU in part as a result of this partition (142). Proteobacteria often have multiple Form I and II rubiscos in their genomes, with different biochemical characteristics, perhaps as a response to their mixotrophic lifestyles (70). Although the chloroplasts of green and non-green algae derive from same endosymbiotic event, Form ID rubiscos in non-green algae are derived from an evolutionary transfer of a Form IC rubisco to their chloroplast genome (LSU and SSU), possibly from an alphaproteobacteria (143). One exception is in dinoflagellates, which have a Form II rubisco acquired from other alphaproteobacteria (144) in their nuclear genomes. Finally, a group of rubiscos were recently found in the genomes of Myoviridae phage and their hosts, Beckwithbacteria (136). This raises the possibility that viruses act as a vehicle for horizontal gene transfer across domains of life. The study of Form I rubisco evolution has benefitted from a clear root in the tree (Figure 4C), evolutionary intermediates have been inferred through metagenomic sampling (74, 145). One such rubisco from *Promineofilum breve* has been termed a Form I’ rubisco because it lacks a small subunit, though it still has an $L_8$ oligomeric form and catalyzes the standard rubisco reaction, albeit with low specificity for $CO_2$ (74).

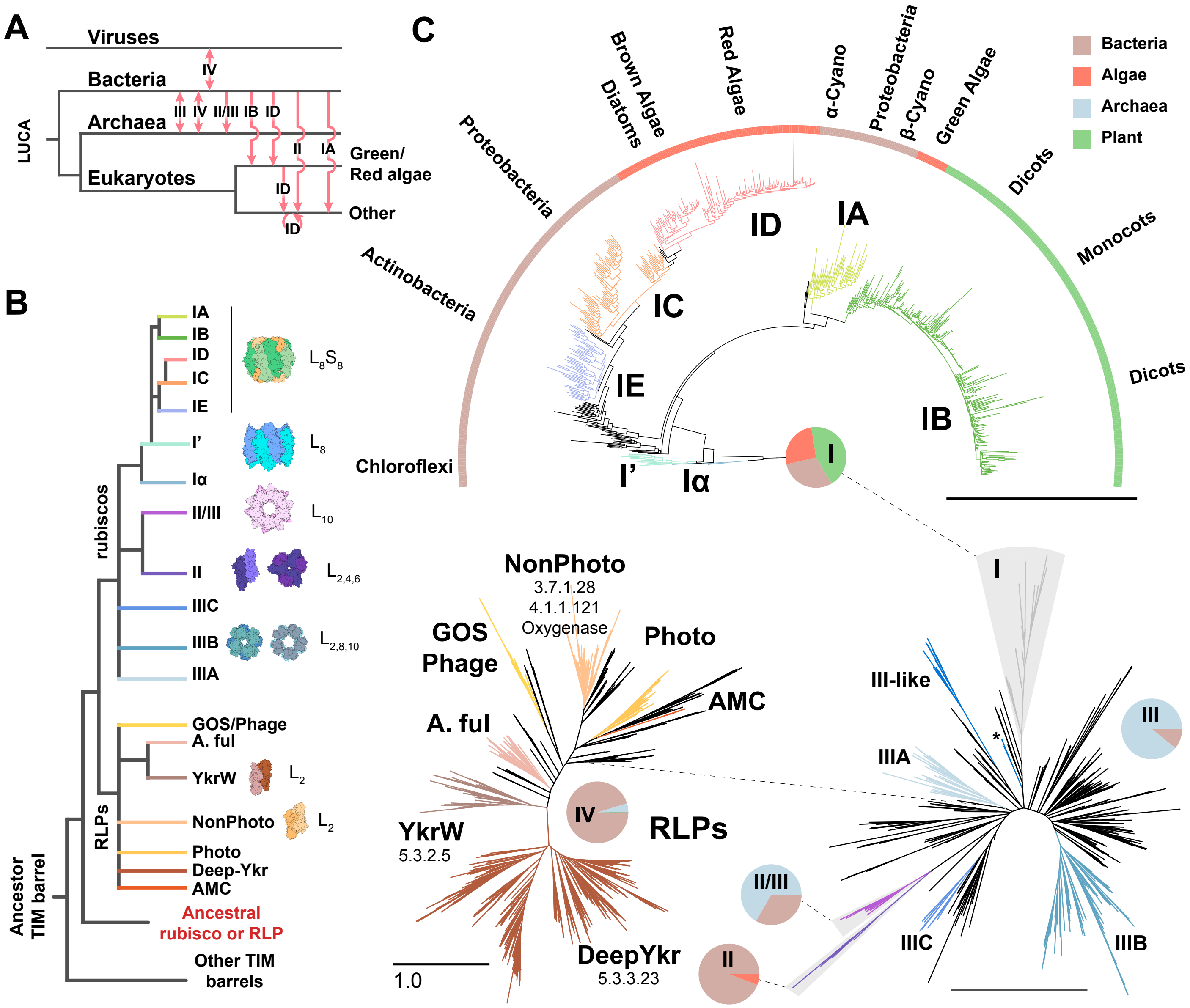

**Figure 4:**
**A)** A schematic adapted from (136) that shows horizontal gene transfer over the history of rubisco evolution. The event labeled IB is the initial endosymbiosis of cyanobacteria to form green algae. The ID event is the transfer of proteobacterial rubisco into the red algal chloroplast (large and small subunits together). Further ID transfers represent secondary or higher level endosymbioses. The Form II transfer is into the dinoflagellate nucleus. Finally the last Form IA represents transfer from a cyanobacteria into *Paulinella*. LUCA is the last universal common ancestor of cellular life. **B)** Simplified diagram of the evolution of the rubisco protein family. The outgroup labeled in red represents the last common ancestor of RLPs and rubisco. It remains an open question what reaction(s) that ancestor catalyzed. Branch lengths in this tree are meaningless. Representative crystal structures of oligomeric complexes are displayed for each clade where they are available (for PDB accessions see Table S3). **C)** A tree of RLPs (Bottom left), all rubiscos (center) and Form I rubiscos (top right). Host organisms are indicated over the Form I tree but are not exclusive. Pie charts and the color bar over the Form I tree represent proportions of clades from a 90% identity dereplication clustering and are colored according to the legend at the top right. The asterisk indicated the location of the transaldolase Form III rubisco variants from (135). Scale bars represent the number of substitutions per site.

## Phylogenetics, trade-offs, and optimization

There have been a number of attempts to capture the natural biochemical diversity of rubiscos in order to better constrain the range of possible kinetic parameters available to the enzyme family. In order to find more divergent rubisco sequences, metagenomic libraries have been assembled and then cloned into bacterial strains that have rubisco knocked out. Metagenomic libraries can come from sources like soil, rivers (146), hydrothermal vents (147), ocean water (148) or deep groundwater stimulated with acetate to promote bacterial growth (146) and the bacterial strains can be photoheterotrophs like *R. capsulatus* (146, 148) or simply *E. coli* (147) - from which metagenomic rubiscos can be purified for characterization. Rubiscos from these samples typically perform worse in *in vivo* complementation or *in vitro* than known rubisco genes, but vast new regions of sequence space remain unexplored, and several new rubisco clades have recently been discovered (e.g. Form III-like, Form Iα (136, 145)). Perhaps the most compelling instance of this approach was a recent survey of Form II rubisco diversity, which uncovered the “fastest” rubisco yet from a betaproteobacterial *Gallionella* species (109), i.e. the rubisco with the highest $k_{cat,C}$ measured thus far. Cyanobacterial rubiscos have generally been assumed to be the fastest variants, so this result suggests there may be additional untapped diversity in natural sequences.

Rubisco is a slowly-evolving enzyme (142) so it is possible that the observed biochemical limitations are a function of phylogenetic constraint - rubiscos may be hampered by their ancestry and not underlying biochemical constraints (105). This interpretation has been called into question (106) since it is clear that in the case of rubiscos with CCMs evolution is at least fast enough to adjust rubisco kinetics to match local $CO_2$ concentration (6). Further sampling and characterization promises to constrain the strength of this “phylogenetic signal” (105). This may help determine to what degree plant rubiscos are evolving too slowly in a changing environment. Metagenomic sampling also promises to reveal additional evolutionary steps and to expand the known functional repertoire of the rubisco superfamily (many Form III rubiscos and RLPs are not found in defined groups).

# Engineering

The motivation for mutagenizing rubisco has been to both test mechanistic hypotheses and improve its function in order to enhance photosynthesis. Point mutations are the smallest steps that can be taken in sequence space and, for many years, were the only mutations accessible. Point mutations near the active site were instrumental in uncovering mechanistic steps (86), but have not been successful in generating enhanced enzymes. Mutations have been installed in many key locations on the enzyme, including the active site and loop 6, without much improvement (17). This is not surprising as it is exceedingly difficult to predict the effect of a mutation on an enzyme’s efficiency, and it is unlikely that a point mutation would have a strong positive effect on catalysis, since evolution would be very likely to find such a modification. Some improvements have been reported, but they are generally weak or else come with a trade-off, for instance, improved $S_{C/O}$ at the cost of a reduction in $k_{cat,C}$ (149).

## Laboratory evolution, trade-offs, and optimization

Directed evolution is the intentional improvement of biological fitness through iterative selection (Figure 5). It is inherently a test of trade-offs and also a tool to explore the biochemical mechanism. There are two components to any experiment in molecular evolution: a library of variants (Figure 5A) and a selection or assay system (Figure 5B). The construction of a library must integrate the goal of the experiment (e.g. is there a particular part of the protein to target?) with the constraints of the assay (e.g. how many variants can be tested?) (150). Libraries can be made in both random and non-random fashions, such as through error-prone

PCR or cloning with designed oligonucleotides, respectively. At ~1500 bases, the gene for the large subunit of rubisco presents many challenges relative to model proteins in that the inherent combinatorics leads to large libraries, complicating cloning, and more difficult assays when using modern tools such as short read DNA sequencing. Recent advances in machine learning, drawing on both sequence and structure, have improved the design of libraries for better exploration of the sequence-function landscape (151, 152) but these have not yet been applied to rubisco.

Selection systems must be developed individually to match the desired protein function. For rubisco, the desired outcome of protein evolution is generally faster $k_{cat,C}$, lower $K_C$ or higher $S_{C/O}$ (ideally all three). Two main modes of selection systems have been developed - genetically tractable autotrophs with replaceable rubisco genes and rubisco dependent *E. coli* (RDE) mutants (Figure 5B). In the case of autotrophs, their growth is rubisco-limited under conditions where they have sufficient energetic resources. Species which have been used to test rubisco replacements include oxidative photoautotrophs like the model cyanobacteria *Synechocystis* sp. PCC6803 (18), as well as facultative autotrophs like *R. capsulatus* (22, 153), *R. palustris* (154, 155) or *C. necator* (156).

The first selection system achieved in *E. coli* relied on rubisco alleviating toxicity induced by *prk* expression (157). RuBP toxicity is a result of it being a metabolic dead-end. No enzyme in *E. coli* makes use of this metabolite, so it shunts away useful carbon and builds up in the cell. Rubisco takes RuBP as a substrate and can alleviate this toxicity. This strategy is inherently prone to "cheaters" which disable *prk* function through mutations. An improved strategy was developed where the *prk* gene is fused to an antibiotic resistance gene, preventing most cheater mutations (158). Implementation in an *E. coli* strain that is especially susceptible to *prk* toxicity enhanced the system further (159).

Other selection systems in *E. coli* are built on the central placement of 3PG, the product of rubisco catalysis; 3PG sits at the nexus of four central metabolic pathways (Figure S3). The first strategy proposed involved a deletion of the GAPDH gene in (8, 19). This metabolic lesion only allows *E. coli* to grow when provided with two carbon sources: one for sugar synthesis and upper glycolysis and another for lipid and energetic pathways including the TCA cycle. GAPDH is a critical link between these two parts of *E. coli* metabolism which has no redundant replacement in the genome. It can, however, be replaced by the combined actions of *prk* and rubisco, which form a link between upper and lower metabolism through the pentose phosphate pathway. *E. coli* with GAPDH removed and *prk*/rubisco added can, in principle, survive on one carbon source alone that would trickle down to the lower metabolism through rubisco, permitting growth. In practice, *E. coli* could not be grown in this mode and required a small amount of added amino acids (160); the resulting selection worked in part because of the same principle of RuBP detoxification. An algorithm was developed to automate the generation of selective strains (127). One such strain was a phosphoglycerate mutase (*gpm)* knockout that was used to evolve an *E. coli* strain capable of growth using rubisco, first to supply all of the upper metabolites in the cell (127) and eventually all carbon biomass (126). Likewise, a ribose-5-phosphate isomerase (*rpi)* knockout was used to investigate the function of $CO_2$ concentrating mechanisms (112). Neither of these latter strains have yet been used to assay rubisco libraries, though additional selection systems have already been proposed (161).

Many different starting points have been chosen for these selections including forms I, II, II/III and III rubiscos (Figure 5A). Because it is difficult to distinguish between improved kinetics and increased expression, the majority of hits from these selections have been rubisco mutants with higher expression levels in the host, often as a result of improved folding and stability. Improvements in kinetic parameters have been modest with a few exceptions. A two-fold increase in $k_{cat}$ was achieved in the Form II/III rubisco from the extremophile *M. burtonii*, though this enzyme had a very low starting rate, implying that it was not optimized for catalysis at room temperature (20). In two cases, a two-fold improvement in $k_{cat}$ was achieved by changing the protein chaperone complement in *E. coli* (17, 21). Due to inconsistencies in the reported rates of rubiscos between different laboratories, it can be difficult to assess or explain these improvements. The selected mutations were all far from the active site so the mechanistic basis for these biochemical improvements remains to be explained. Additional kinetic measurements and structural studies are necessary.

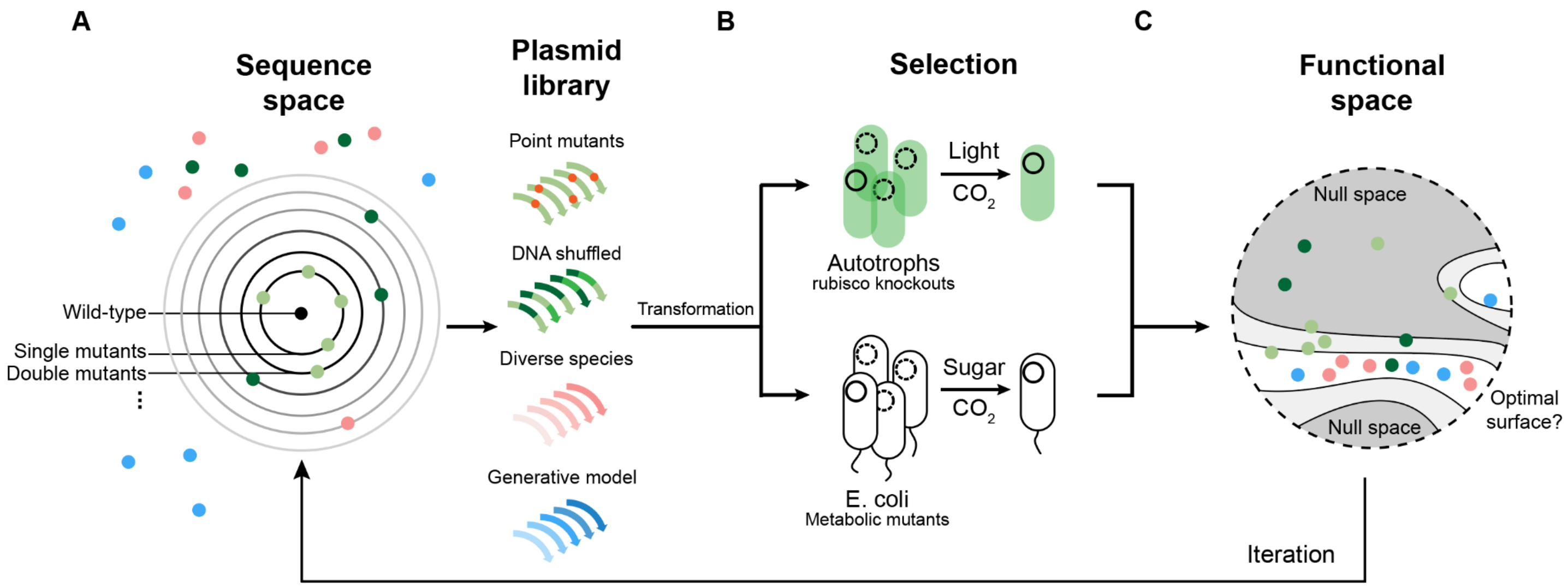

**Figure 5:**
A) Mutant libraries are made in a variety of ways resulting in different distributions throughout sequence space. Libraries are made through environmental sampling, random mutagenesis, or synthesized oligo pools. Distance in sequence space is measured in number of mutations. Sequence space is discrete but high-dimensional. **B)** Genetic screening for functional rubisco can be accomplished using either rubisco knockout strains of autotrophs, or rubisco-dependent mutants of *E. coli*. **C)** Mutants that survive selection may have clustered properties. Analysis of these clusters may reveal optimal surfaces in functional space. Functional space is also high-dimensional. Dimensions include, but are not limited to: $k_{cat,C}$, $k_{cat,,O}$, $K_C$, $K_O$, $T_m$, $K_{M,RuBP}$, $k_3$, $k_6$, $k_5$, $k_8$, $K_E$ (the enolization constant), turnover number, maximum expression level, average activation level, $k_{on,RuBP}$, $k_{off,3PG}$, $k_{off,2PG}$, etc.

# Plant Synthetic Biology

One major goal in rubisco engineering is to improve plant growth. As atmospheric $CO_2$ levels climb, it may be possible to improve plant growth by simply replacing the rubisco genes of crop plants with enzymes that are adapted to higher $CO_2$ levels like those of cyanobacteria. It has been possible to edit the chloroplast genome of the model plant, *N. tabacum*, since 1993 (162) and soon thereafter a transgenic line with the rubisco large subunit moved to the nucleus was created (163). Since then, a variety of rubiscos of different forms have been expressed in the tobacco chloroplast, in some cases with specific chaperones (Table S4). These mutant tobacco plants grow uniformly slower than wild-type, even in chambers with $CO_2$ elevated to 1%. They also have higher $CO_2$ compensation points (i.e. the $CO_2$ concentration where there is net assimilation of carbon), in many cases because the rubiscos used have lower $CO_2$ affinity than the endogenous tobacco gene. In nearly all cases, the cause of poor growth is clear as rubisco expression levels are reduced by an order of magnitude compared to WT (164). In one case, rubisco levels were only reduced by a factor of two and growth at high $CO_2$ was comparable to WT (165). Likewise, extremely similar rubiscos have been swapped with minimal effect (166). Although the Whitney lab has developed a tobacco line with *R. rubrum* rubisco which is more readily transformable (167), chloroplast transformation remains an enormous bottleneck and all such studies suffer from the ability to test few rubisco variants *in planta*.

# Perspectives and future work

Despite decades of research, the promises of rubisco engineering remain unrealized at every level. There are still deep unanswered questions about the chemical mechanism of catalysis and what implications it has on the prospects for improving rubisco carboxylation rate and specificity. The evolutionary constraints on the enzyme are unknown and poorly sampled. Massively improved rubisco variants have not emerged, and finally, improvements in plant photosynthesis have not been achieved through rubisco replacement. On the other hand, much progress has been made in revealing the chemical mechanism of oxygenation (84, 91). Metagenomic mining has provided an enormous wealth of rubisco isoforms to study mechanism and evolution (136). High throughput biochemistry has recently uncovered a record-breaking rubisco variant (109), and improvements in plant growth have been achieved through increasing rubisco levels (168, 169).

One source of future advances may be from additional detailed biochemical characterization of phylogenetically distant rubisco isoforms which will better constrain the full breadth of rubisco catalytic possibilities. This knowledge will come in the form of a sharper account of the etiology of trade-offs. An added benefit of characterizations of previously overlooked rubisco isoforms is a new source of starting points for directed evolution. The specificity of plant chaperones for their cognate rubiscos also remains a large barrier to heterologously assaying and engineering variants, and deciphering this "code" will surely accelerate advances on Form I.

Continued improvements in selection systems (Figure S3) are essential in order to test larger libraries which would unlock the benefits of computational design approaches. Advances in machine learning models, coupled with mounting empirical sequences and structural data, will also help explore the functional regions of the sequence landscape, which is vast due to the large size of rubisco. For example, many bacterial rubiscos apparently do not have the same chaperone requirements as relatively similar plant variants. Machine learning methods (151, 152) may be able to re-combine sequences in a productive fashion in order to better understand folding landscapes and to isolate variants combining the most desirable features of Form I enzymes. Alternatively, even if future experimentation shows that rubisco is constrained by trade-offs, laboratory evolution promises to reveal the precise shape of those constraints (optimal surface in Figure 5C) and to test hypotheses in a proof-by-synthesis fashion. For example, current trade-off models suggest that rubiscos with higher $k_{cat}$s and lower specificities should be evolvable from slower and more specific variants.

It is likely that the complexity of expressing rubisco *in planta* has limited experimental successes, and so improvements in plant genome editing will be key to translating rubisco improvements into the field. Successful heterologous expression of rubisco *in planta* depends on the ability to test a large number of alternative gene cassettes in order to balance gene expression of large and small subunits along with all of their associated chaperones. This is especially true for Form IB and ID rubiscos that have a heavy dependence on the proteostatic machinery for folding and activation (82, 170, 171). Of particular importance is further improving chloroplast genome editing technology, which currently suffers from low throughput and long time horizons. Even if biochemical improvements fail to materialize, heterologous replacement of rubisco from red algae is predicted to be advantageous (171, 172), due to the slow evolution of rubisco in plants (142).

If direct improvements to rubisco fail to materialize, alternative strategies for photosynthetic engineering, around the constraints of rubisco, have been proposed and are being pursued (173). There are several successes in the engineering of pathways closely related to extant plant physiology, such as in the case of the photorespiratory bypass (174) or increased expression of CBB components (175), which led to double-digit percent yield increases. Alternatively, there are more ambitious proposals for re-engineering plant physiology with CCM components that could offer greater gains but that come with the downside of significant complexity (111, 112, 176). Looking further into the future, it may eventually be possible to circumvent the downsides of rubisco entirely by switching to alternative carbon fixation cycles, either natural or artificial. Until then, rubisco engineering will hold a key position in efforts to improve photosynthesis.

# Acknowledgements:

We thank Jacob West-Roberts, Alex Jaffe, Jill Banfield and Nat Thompson for helpful conversations in building rubisco trees. We thank Jack Desmarais and Elad Noor for invaluable advice related to thermodynamic interpretations and help with data analysis. We thank Abhishek Bhatt for help in assembling the carboxylase table. We thank Avi Flamholz as well as Luke Oltrogge and Muntathar Al-Shimary for their insightful comments when preparing the manuscript. The work was supported by the Howard Hughes Medical Institute (HHMI), a K99 award 5K99GM141455-02 to N.P., National Science Foundation grant (MCB-1818377) to D.F.S. and US Department of Energy grant DE-SC00016240 to D.F.S.